\documentclass[onecolumn,11pt,journal]{IEEEtran}

\usepackage[letterpaper,margin=1in]{geometry}

\usepackage{amsmath,amssymb,amsfonts,bm}
\usepackage{bbm}
\usepackage{graphicx}
\usepackage{array}
\usepackage{booktabs}
\usepackage{cite}
\usepackage{url}
\usepackage{xcolor}
\usepackage[caption=false,font=footnotesize]{subfig}

\usepackage{algorithm}
\usepackage{algpseudocode}

\usepackage{tikz}
\usetikzlibrary{arrows.meta,positioning,shapes.geometric,calc,fit,backgrounds}

\pgfdeclarelayer{background}
\pgfsetlayers{background,main}

\usepackage{pdfrender}

\tikzset{
  flow/.style={
    draw=red!80!black, very thick,
    rounded corners=2mm,
    inner sep=2.6pt,
    align=center,
    fill=white
  },
  flowwide/.style={flow, text width=0.83\columnwidth},
  flowformula/.style={flowwide, font=\scriptsize},
  term/.style={flow, ellipse, minimum width=1.65cm, minimum height=6mm},
  decision/.style={
    draw=red!80!black, very thick,
    diamond, aspect=2.15,
    inner sep=1.6pt,
    align=center,
    fill=white
  },
  dblarr/.style={
    double, line width=0.5pt,
    -{Stealth[length=2.2mm]},
    draw=red!80!black
  },
  looparr/.style={dblarr, rounded corners},
  lab/.style={font=\scriptsize, inner sep=1pt},
  grouplab/.style={lab, fill=white, inner sep=1.4pt}
}
\newcommand{\convpanelmark}[2][4mm]{%
  \par\vspace{1pt}%
  {\footnotesize\normalfont\mdseries
   \hbox to \linewidth{%
     \hfil\kern#1 (#2)\kern-#1\hfil
   }}%
}
\usepackage[final]{microtype}

\begin{document}

\title{Rapidly Convergent Finite-Element Domain Decomposition Method With Two-Channel Transmission Conditions}

\author{Furkan~\c{S}\i k,~\IEEEmembership{Graduate Student Member,~IEEE,}
     Fernando~L.~Teixeira,~\IEEEmembership{Fellow,~IEEE,}
and~Balasubramaniam~Shanker,~\IEEEmembership{Fellow,~IEEE}
\thanks{Manuscript received MM DD, YYYY; revised MM DD, YYYY.}
\thanks{This work was supported in part by the Ohio State 
President's Research Excellence Program and by the 
Ohio Supercomputer Center Grant PAS-0061.}
\thanks{The authors are with the Department of Electrical and Computer Engineering and ElectroScience Laboratory, The Ohio State University, Columbus, OH 43210, USA (\textit{Corresponding author: Furkan \c{S}\i k})}}

\maketitle
\raggedbottom

\begin{abstract}
A novel dual-primal finite element tearing and interconnecting (FETI-DP) domain decomposition method (DDM) is introduced for solving Maxwell's equations. The proposed method is built upon a two-channel transmission condition (TC) enforcing both tangential-field and normal-flux continuity across subdomain interfaces. Two interface channels are separately constructed from the Faraday and Amp\`ere-Maxwell equations. Simultaneously enforcing tangential-field and normal-flux continuity reduces interface jumps by several orders of magnitude relative to the widely used Robin TC. The resulting global interface equation exhibits markedly improved iterative convergence. Numerical experiments across different partitioning strategies, subdomain counts and volumes, mesh resolutions, and geometries  demonstrate drastic reductions in iteration counts, while accuracy, massive parallelism, and scalability are preserved.
\end{abstract}

\section{Introduction}

\IEEEPARstart{D}{espite} its inherent advantages and widespread adoption for solving electromagnetic (EM) boundary value problems governed by Maxwell's equations, the finite element method (FEM) produces, for electrically large problems, sparse linear systems with millions or even billions of unknowns, straining both direct and iterative solvers~\cite{reddy2024computational,peterson1998computational,Zienkiewicz2005FEM,jin2015finite,ming1lam,toselli2005domain,zhao2009domain}. Among non-overlapping domain decomposition methods (DDMs), the finite element tearing and interconnecting (FETI) family addresses this challenge by tearing the computational domain into independently factorized subdomains and connecting them through interface Lagrange multipliers, thereby reducing a three-dimensional (3-D) volumetric problem to a 2-D global interface equation solved iteratively by Krylov subspace methods~\cite{farhat1991feti,ming1lam,ming2lam,farhat2005fetidp,farhat2001fetidp,jin2015finite,toselli2005domain,roux2009feti2lm,zhao2009domain,paraschos2012robust}. In the dual-primal (FETI-DP) variants, corner-edge unknowns remain primal and furnish a global coarse problem~\cite{ming1lam,ming2lam,farhat2005fetidp,farhat2001fetidp}. 

The transmission condition (TC) imposed at the subdomain interfaces governs both the well-posedness of the local subdomain problems and the iterative convergence of the global interface equation. Robin TC, imposed with two multipliers per interface (FETI-DP-$2\lambda$), has become the de facto standard~\cite{jin2015finite,ming2lam,paraschos2012robust}. However, the Robin TC damps only the propagating spectrum of the interface error modes. The evanescent modes are beyond its reach: no single Robin coefficient can damp both TE- and TM-type evanescent families simultaneously~\cite{peng2010oneway,dolean2015effective,mingsecond,peng2010nonconformal}. When applied to challenging problems with large subdomain counts, many corner edges, resonances, multiscale features, and deep-evanescent waves, convergence slows down and even stagnates. Second-order TCs (SOTCs) target evanescent modes with tangential-derivative augmentations, but at a price: their per-interface auxiliary currents and charges become redundant on corner edges, rendering the DDM matrix nearly singular and necessitating dedicated corner edge penalties, while for deep-evanescent waves the damping deteriorates and reverts toward Robin behavior~\cite{luconverge,lucorner,peng2010nonconformal,rawat2010nonoverlapping}. Consequently, none of these TCs robustly guarantees convergence on large-scale EM problems. This shortfall has been driving practitioners toward direct-solver-based DDMs, which avoid convergence stagnation altogether. However, direct solvers sacrifice several defining strengths of DDM, namely massive parallelism, computational scalability, and computational efficiency~\cite{moshfegh2019parallel,directddmmarin,moshfegh2016d3m}. The open problem, therefore, is to devise a FETI-DDM whose global interface equation remains inexpensive to solve with robust iterative solvers, exhibits rapid stagnation-free convergence across diverse problem settings, and preserves massive parallelism and computational scalability.

This letter introduces such a method, built up on a two-channel TC constructed directly from the mixed $\mathbf{E}$-$\mathbf{B}$ FE formulation~\cite{mixedEB,HeTAP} using Whitney edge and face elements on conformal meshes. Their two-sided enforcement yields four Lagrange multipliers per interface while retaining dual-primal corner treatment. The Faraday channel pairs $\hat{\bm n}\times\mathbf{E}$ with $\hat{\bm n}(\hat{\bm n}\cdot\mathbf{B})$, while the Amp\`ere-Maxwell channel pairs $\hat{\bm n}\times\mathbf{H}$ with $\hat{\bm n}(\hat{\bm n}\cdot\mathbf{D})$. 
The assignment of each evanescent family its own channel removes the single-coefficient obstruction of the Robin TC, with monotone damping that persists in the deep-evanescent regime (where the Robin TC stagnates and the SOTCs rebound~\cite{peng2010nonconformal,peng2010oneway}).
The proposed TC enforces tangential-field and normal-flux continuity simultaneously, reducing interface jumps of $\hat{\bm n}\times\mathbf{E}$ and $\hat{\bm n}\cdot\mathbf{B}$ by orders of magnitude relative to the Robin TC. The restored continuity accelerates convergence of the global interface equation and improves accuracy, particularly at relaxed solver tolerances.

Extensive numerical experiments across waveguide geometries, partitionings, discretizations, and subdomain counts and sizes, including problems with multiscale features and deep-evanescent wave content, validate the proposed method. With no loss of accuracy, GMRES iteration counts drop drastically relative to the widely used Robin TC and SOTC-based approaches.

\section{Formulation}

This section derives the two-channel TC from the first-order curl equations and assembles the resulting FETI-DP-$4\lambda$ system. Let $\Omega=\bigcup_{i=1}^{N_s}\Omega^i$ be a nonoverlapping conformal partition of the domain of interest, with interfaces $\Gamma^{ij}=\partial\Omega^i\cap\partial\Omega^j$, neighbor sets $\mathcal N^i=\{j\ne i:\Gamma^{ij}\ne\varnothing\}$, local interfaces $\Gamma^i=\bigcup_{j\in\mathcal N^i}\Gamma^{ij}$, and outward unit normals satisfying $\hat{\bm n}^{i}=-\hat{\bm n}^{j}$ on $\Gamma^{ij}$; edges incident in more than two subdomains form the corner edge set $\mathcal E_c$. The fields in $\Omega^i$ satisfy the Faraday and Amp\`ere-Maxwell equations,
\begin{align}
 \nabla\!\times\mathbf E^i&=-j\omega\mathbf B^i,
 \label{eq:faraday}\\
 \nabla\!\times\mathbf H^i&=\mathbf J^i+j\omega\mathbf D^i,
 \label{eq:ampere}
\end{align}
with $\mathbf D^i=\varepsilon_i\mathbf E^i$, $\mathbf H^i=\mu_i^{-1}\mathbf B^i$, and impressed electric current density $\mathbf J^i$. Boldface denotes physical vector fields, an overbar a discrete coefficient vector, and a double overbar a matrix. The superscripts $i$ and $(i)$ index the subdomain number in continuous and discrete variables, respectively, and the superscript $S$ denotes a Schur-reduced quantity.

Rather than reducing \eqref{eq:faraday}-\eqref{eq:ampere} to a second-order vector wave equation, we retain $\mathbf E^i\in W^1(\Omega^i)\subset H(\operatorname{curl};\Omega^i)$ and $\mathbf B^i\in W^2(\Omega^i)\subset H(\operatorname{div};\Omega^i)$, expanded in the Whitney edge and face bases $\mathbf W_a^{(1)}$ and $\mathbf W_f^{(2)}$~\cite{HeTAP,crawfordbs}, collecting the $N_e^{(i)}$ edge and $N_f^{(i)}$ face coefficients in $\bar e^{(i)}$ and $\bar b^{(i)}$; the mixed subdomain unknown is ordered as $\bar x^{(i)}=(\bar b^{(i),T},\bar e^{(i),T})^T$. Testing \eqref{eq:faraday} against $\mu_i^{-1}\mathbf W_f^{(2)}$ and \eqref{eq:ampere} against $\mathbf W_a^{(1)}$ yields a local block system whose interface data are the two tangential pairings
\begin{equation}
 \langle\mu_i^{-1}\mathbf W_f^{(2)},
 \hat{\bm n}^i\!\times\mathbf E^i\rangle_{\Gamma^i},\qquad
 -\langle\mathbf W_a^{(1)},
 \hat{\bm n}^i\!\times\mathbf H^i\rangle_{\Gamma^i},
 \label{eq:interface_pairings}
\end{equation}
entering the face and edge rows, respectively; here $\langle\mathbf u,\mathbf v\rangle_{\Omega^i}=\int_{\Omega^i}\mathbf u\cdot\mathbf v\,\mathrm{d}\Omega$ and $\langle\mathbf u,\mathbf v\rangle_{\Gamma^i}=\int_{\Gamma^i}\mathbf u\cdot\mathbf v\,\mathrm{d}\Gamma$.

The proposed two-channel TC supplies these pairings by assigning one channel to each curl equation. On the side $\Gamma^{ij}\subset\partial\Omega^i$, the Faraday and Amp\`ere-Maxwell channel traces of $\Omega^i$, built from its own fields, are:
\begin{align}
 \bm\Lambda_1^{ij}&=\hat{\bm n}^{i}\!\times\mathbf E^i
 +a_{B,i}\,\hat{\bm n}^{i}\big(\hat{\bm n}^{i}\!\cdot\mathbf B^i\big),
 \label{eq:faraday_channel}\\
 \bm\Lambda_2^{ij}&=\hat{\bm n}^{i}\!\times\mathbf H^i
 +a_{D,i}\,\hat{\bm n}^{i}\big(\hat{\bm n}^{i}\!\cdot\mathbf D^i\big).
 \label{eq:ampere_channel}
\end{align}
The tangential traces carry unit weight, while the flux coefficients $a_{B,i}$ and $a_{D,i}$ carry the local wave speed $v_i=1/\sqrt{\mu_i\varepsilon_i}$, rendering the two terms of each channel commensurate; they are set following the Fourier analysis used to design SOTCs and analyze Robin TCs~\cite{lee2005nonoverlapping,peng2010oneway,peng2010nonconformal}. Each side also carries two independent Lagrange multipliers (dual variables) $\bm\lambda_1^{ij},\bm\lambda_2^{ij}$ (coefficient vectors $\bar\lambda_1^{ij},\bar\lambda_2^{ij}$), acting as incoming interface data through the boundary condition imposed on $\Omega^i$ along $\Gamma^{ij}$,
\begin{equation}
 \bm\Lambda_q^{ij}+\bm\lambda_q^{ij}=\bm 0,\qquad q=1,2.
 \label{eq:local_closure}
\end{equation}
The multiplier vectors of all interfaces of $\Omega^i$ are collected into $\bar\lambda^{(i)}$; with two channels per side and two sides per interface, each interface carries four Lagrange multipliers. Inserting \eqref{eq:faraday_channel}-\eqref{eq:ampere_channel} into \eqref{eq:local_closure} and solving for the tangential traces on each side $\Gamma^{ij}$ gives
\begin{align}
 \hat{\bm n}^{i}\!\times\mathbf E^i&=-\bm\lambda_1^{ij}
 -a_{B,i}\,\hat{\bm n}^{i}\big(\hat{\bm n}^{i}\!\cdot\mathbf B^i\big),
 \label{eq:faraday_trace}\\
 \hat{\bm n}^{i}\!\times\mathbf H^i&=-\bm\lambda_2^{ij}
 -a_{D,i}\,\hat{\bm n}^{i}\big(\hat{\bm n}^{i}\!\cdot\mathbf D^i\big),
 \label{eq:ampere_trace}
\end{align}
and substituting them into \eqref{eq:interface_pairings}, summed over $j\in\mathcal N^i$, yields the local problem with the TCs imposed:
\begin{equation}
 \bar{\bar A}^{(i)}\bar x^{(i)}+\bar{\bar G}^{(i)}\bar\lambda^{(i)}=\bar f^{(i)},
 \label{eq:local_mixed_problem}
\end{equation}
\begin{equation}
 \bar{\bar A}^{(i)}=
 \begin{bmatrix}
  j\omega\bar{\bar M}_B^{(i)}-a_{B,i}\bar{\bar N}_B^{(i)}&\bar{\bar C}^{(i),\mathrm a}\\
  -\big(\bar{\bar C}^{(i)}\big)^{T}&j\omega\bar{\bar M}_E^{(i)}+a_{D,i}\bar{\bar N}_D^{(i)}
 \end{bmatrix},
 \label{eq:local_mixed_matrix}
\end{equation}
\begin{equation}
 \bar{\bar G}^{(i)}=\operatorname{blkdiag}_{j\in\mathcal N^i}
 \big(-\bar{\bar Q}_{1}^{ij},\;\bar{\bar Q}_{2}^{ij}\big).
 \label{eq:channel_load}
\end{equation}
Here $[\bar{\bar M}_B^{(i)}]_{ff'}=\langle\mu_i^{-1}\mathbf W_f^{(2)},\mathbf W_{f'}^{(2)}\rangle_{\Omega^i}$ and $[\bar{\bar M}_E^{(i)}]_{aa'}=\langle\varepsilon_i\mathbf W_a^{(1)},\mathbf W_{a'}^{(1)}\rangle_{\Omega^i}$ are the face and edge mass matrices;
$[\bar{\bar C}^{(i)}]_{fa}
=\langle\mu_i^{-1}\mathbf W_f^{(2)},
\nabla\!\times\mathbf W_a^{(1)}\rangle_{\Omega^i}$
is the discrete curl coupling, and $\bar{\bar C}^{(i),\mathrm a}$ its adjoint block, obtained by moving the Faraday curl onto the test functions to expose the first pairing of \eqref{eq:interface_pairings} and differing from $\bar{\bar C}^{(i)}$ by the interface trace of $\hat{\bm n}^i\!\times\mathbf W_a^{(1)}$; $\bar f^{(i)}$ is the tested impressed current and exterior boundary data. The multiplier pairings, with $\bm\lambda_1^{ij},\bm\lambda_2^{ij}$ expanded in interface traces of $W^2$ and $W^1$, form interface pairing blocks $\bar{\bar Q}_{1}^{ij}$ (face rows) and $\bar{\bar Q}_{2}^{ij}$ (edge rows) of the channel load $\bar{\bar G}^{(i)}$. The flux pairings assemble into the normal operators: $\bar{\bar N}_B^{(i)}$ is the Gram matrix of the $W^2$ normal traces, which the mixed $\mathbf E$-$\mathbf B$ setting carries directly on $\Gamma^i$, $\bar{\bar N}_D^{(i)}$ realizes the electric normal flux from the tangential electric trace as a surface grad-div operator. Assembled per interface, $\bar{\bar N}_B^{(i)}$ and $\bar{\bar N}_D^{(i)}$ are identical on the two sides, and neither introduces an auxiliary interface unknown.

The two sides are interconnected by the neighbor-side counterpart of \eqref{eq:local_closure}: $\bm\lambda_q^{ij}+\bm\Lambda_q^{ij}[\mathbf E^j,\mathbf B^j]=\bm 0$ on $\Gamma^{ij}$, where $\bm\Lambda_q^{ij}[\mathbf E^j,\mathbf B^j]$ is the trace \eqref{eq:faraday_channel}-\eqref{eq:ampere_channel} built from the fields of $\Omega^j$ with normal $\hat{\bm n}^i$. Together with \eqref{eq:local_closure} on $\Omega^i$, this enforces $\bm\Lambda_q^{ij}[\mathbf E^i,\mathbf B^i]=\bm\Lambda_q^{ij}[\mathbf E^j,\mathbf B^j]$ at convergence. Under $\hat{\bm n}^{i}=-\hat{\bm n}^{j}$, eliminating the neighbor's tangential trace with \eqref{eq:local_closure} on $\Omega^j$ leaves the discrete exchange
\begin{equation}
 \bar\lambda_q^{ij}+\bar{\bar P}_q^{ij}\big(\bar\lambda_q^{ji}
 +\bar{\bar T}_q^{ji}\,\bar x^{(j)}\big)=\bar 0,
 \qquad q=1,2,
 \label{eq:channel_exchange}
\end{equation}
with $\bar{\bar P}_q^{ij}$ the signed edge/face orientation matrix expressing traces on $\Gamma^{ji}$ in the orientation of $\Gamma^{ij}$, and $\bar{\bar T}_q^{ji}$ the flux pairing rows restricted to $\Gamma^{ji}$.

Let $\bar x_r$ collect the non-corner (remainder) coefficients of all subdomains and $\bar x_c$ the globally assembled primal corner edge coefficients on $\mathcal E_c$; subscripts $r$ and $c$ label this partition throughout. Partitioning \eqref{eq:local_mixed_problem} accordingly, $\bar{\bar K}_{rr}=\operatorname{blkdiag}_i\bar{\bar A}_{rr}^{(i)}$ and $\bar{\bar K}_{rc}$, $\bar{\bar K}_{cr}$, $\bar{\bar K}_{cc}$ carry the corner couplings of $\bar{\bar A}^{(i)}$, while $\bar{\bar G}_r$, $\bar{\bar G}_c$ collect the $r$/$c$ rows of the channel loads $\bar{\bar G}^{(i)}$. The interconnection rows close the system: $\bar{\bar T}_r$, $\bar{\bar T}_c$ collect the reoriented flux pairings $\bar{\bar P}_q^{ij}\bar{\bar T}_q^{ji}$ of \eqref{eq:channel_exchange} and $\bar{\bar D}_\lambda$ its multiplier blocks $(\bar{\bar I},\bar{\bar P}_q^{ij})$, giving
\begin{equation}
 \begin{bmatrix}
  \bar{\bar K}_{rr}&\bar{\bar K}_{rc}&\bar{\bar G}_r\\
  \bar{\bar K}_{cr}&\bar{\bar K}_{cc}&\bar{\bar G}_c\\
  \bar{\bar T}_r&\bar{\bar T}_c&\bar{\bar D}_\lambda
 \end{bmatrix}
 \begin{bmatrix}\bar x_r\\\bar x_c\\\bar\lambda\end{bmatrix}
 =\begin{bmatrix}\bar f_r\\\bar f_c\\\bar 0\end{bmatrix}.
 \label{eq:feti4_bordered}
\end{equation}
Eliminating $\bar x_r$ with $\bar{\bar Z}=\bar{\bar K}_{rr}^{-1}$, one sparse factorization per subdomain, condenses \eqref{eq:feti4_bordered} onto $(\bar x_c,\bar\lambda)$ with
\begin{equation}
 \begin{alignedat}{2}
  \bar{\bar K}_{cc}^{S}&=\bar{\bar K}_{cc}
   -\bar{\bar K}_{cr}\bar{\bar Z}\bar{\bar K}_{rc},&\quad
  \bar{\bar G}_c^{S}&=\bar{\bar G}_c
   -\bar{\bar K}_{cr}\bar{\bar Z}\bar{\bar G}_r,\\
  \bar{\bar T}_c^{S}&=\bar{\bar T}_c
   -\bar{\bar T}_r\bar{\bar Z}\bar{\bar K}_{rc},&\quad
  \bar{\bar D}_\lambda^{S}&=\bar{\bar D}_\lambda
   -\bar{\bar T}_r\bar{\bar Z}\bar{\bar G}_r,
 \end{alignedat}
 \label{eq:schur_blocks}
\end{equation}
and $\bar f_c^{S}=\bar f_c-\bar{\bar K}_{cr}\bar{\bar Z}\bar f_r$, $\bar g^{S}=-\bar{\bar T}_r\bar{\bar Z}\bar f_r$. Eliminating the primal corner block produces the \textit{global interface equation} of the proposed FETI-DP-$4\lambda$ approach:
\begin{align}
 \bar{\bar F}_{4\lambda}\bar\lambda=\bar d_{4\lambda},\qquad
 \bar{\bar F}_{4\lambda}&=\bar{\bar D}_\lambda^{S}
 -\bar{\bar T}_c^{S}(\bar{\bar K}_{cc}^{S})^{-1}\bar{\bar G}_c^{S},
 \label{eq:f4lambda}\\
 \bar d_{4\lambda}&=\bar g^{S}
 -\bar{\bar T}_c^{S}(\bar{\bar K}_{cc}^{S})^{-1}\bar f_c^{S},
 \label{eq:f4lambda_rhs}
\end{align}
solved by GMRES. Its action involves only subdomain-level solves and one small corner (coarse) solve propagating residual information globally; after convergence, all field and flux coefficients follow by local back-substitution.

Compared with FETI-DP formulations built on the second-order vector wave equation~\cite{ming1lam,ming2lam}, in which the electric field is the only subdomain unknown, the mixed discretization carries both fields, storing $N_e^{(i)}+N_f^{(i)}$ rather than $N_e^{(i)}$ coefficients. The second channel raises the dual trace count from $2N_e^\Gamma$ to $2(N_e^\Gamma+N_f^\Gamma)$. Both ratios are fixed by the mesh: the added subdomain block is a face mass matrix carrying no curl, the added interface blocks are trace Gram matrices and signed permutations. No additional subdomain solves are required. The larger local systems and longer multiplier vector raise the cost of a single iteration by a bounded factor, whereas the iteration counts fall by far larger factors. The gain widens with problem difficulty, precisely where Robin TC slows or stagnates. The proposed TC therefore keeps the global interface equation solvable by a robust iterative solver at a total solution cost well below that of the Robin FETI-DP baseline.
\section{Numerical Experiments}
\label{sec:results}

The proposed two-channel FETI-DP-$4\lambda$ method is validated on a sequence of 3-D examples against the Robin TC~\cite{ming2lam,jin2015finite} and SOTCs~\cite{peng2010oneway,peng2010nonconformal} using identical conformal tetrahedral meshes and solver settings. The subdomain problems are discretized in the mixed $\mathbf{E}$-$\mathbf{B}$ variables of Section~II and factorized by a sparse direct solver, while the global interface equation is solved by unpreconditioned GMRES with a restart parameter of 1200 down to $10^{-10}$ relative residual tolerance. Accuracy is assessed against the Robin TC-based FETI-DP-$2\lambda$ solution and, where available, analytical or semi-analytical references; GMRES convergence and interface field and flux jumps are also examined. All waveguide examples are excited at the input port in the dominant TE$_{10}$ mode; except for the input and output ports, all boundaries are perfect electric conductors.

\subsection{WR-90 Waveguide: Uniform and Irregular Partitionings}
\label{sec:wr90}

We first consider a WR-90 waveguide operating at $f=8.2$~GHz, partitioned into $40$ uniform slabs of width $\Delta x=6$~mm with $39$ cross-section interfaces. Both TCs agree with the analytical solution to approximately $99.25\%$. The relative $L^{2}$-norm difference between the proposed and Robin TC electric-field solutions is only $3.2\times10^{-7}$. As shown in Table~\ref{tab:wr90}, the proposed TC reduces the GMRES iteration count relative to the Robin TC by factors of $3.42$ and $4.42$ at the $10^{-6}$ and $10^{-10}$ residual relative tolerances, respectively. At the $10^{-10}$ tolerance, the corresponding wall-clock time decreases from $380.7$\,s to $189.3$\,s, yielding a $\sim2\times$ speedup.

\begin{figure}
  \centering
  \includegraphics[width=0.8\textwidth]{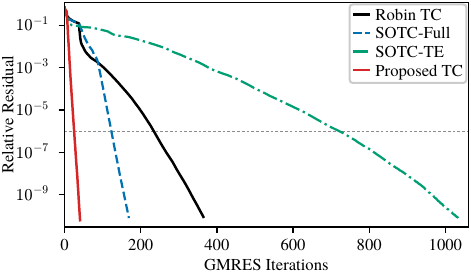}
  \caption{GMRES convergence history of the global interface equation for the irregularly partitioned waveguide.}
  \label{fig:irr}
\end{figure}

To test convergence and robustness of the global interface equation across different subdivisions and subdomain volumes, the same waveguide is partitioned into $30$ irregular slabs with randomly distributed widths $\Delta x\in[0.8,\,18.8]$~mm; the closely spaced thin slabs couple through evanescent modes that the Robin TC leaves undamped and that degrade the SOTC damping. As seen in Fig.~\ref{fig:irr}, the proposed TC converges in almost $9$ times fewer iterations than the Robin TC, while the two solutions agree to a relative $L^{2}$ difference of only $0.0001\%$.

\begin{table}[t]
\caption{GMRES Iteration Counts for the Uniformly Partitioned Waveguide}
\label{tab:wr90}
\centering
\footnotesize
\setlength{\tabcolsep}{6pt}
\begin{tabular}{@{}lcc@{}}
\toprule
& \multicolumn{2}{c}{Tolerance} \\
\cmidrule(lr){2-3}
Transmission Condition & $10^{-6}$ & $10^{-10}$ \\
\midrule
Robin TC    & 188 & 301 \\
SOTC-Full   & 136 & 213 \\
Proposed TC & $\bm{55}$ & $\bm{68}$ \\
\bottomrule
\end{tabular}
\end{table}

\subsection{Ridge-Loaded, L-Shaped, and METIS-Partitioned Waveguides}
\label{sec:wgsuite}
 
\begin{figure}[t]
\centering
\begin{minipage}[b]{0.47\textwidth}
  \centering
  \includegraphics[width=\linewidth]{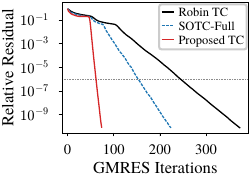}
  \convpanelmark{a}
\end{minipage}\hspace{0.04\textwidth}%
\begin{minipage}[b]{0.47\textwidth}
  \centering
  \includegraphics[width=\linewidth]{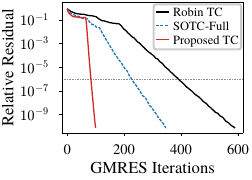}
  \convpanelmark{b}
\end{minipage}
\caption{GMRES convergence history of the global interface equation for the L-shaped waveguide partitioned into (a) $40$ and (b) $100$ subdomains.}
\label{fig:lshape_conv}
\end{figure}

Three further examples introduce a ridge discontinuity, a right-angle bend, and an irregular METIS partitioning. The ridge-loaded waveguide constricts the cross section to half-width over a ridge section and is partitioned into $40$ subdomains. The L-shaped waveguide introduces a right-angle bend and is partitioned into $40$ subdomains, with an additional $100$-subdomain case to probe robustness to subdomain count. The WR-90 waveguide is also partitioned by METIS~\cite{karypis1997metis} into $16$ irregular subdomains with $37$ interfaces and multiple corner edges. At $10^{-10}$ tolerance, the proposed TC reduces the GMRES iteration count by factors of 3.39 and 9.53 relative to the Robin TC for the ridge-loaded and METIS-partitioned examples, respectively, with no loss of accuracy; similar reductions are observed in Fig.~\ref{fig:lshape_conv} for the L-shaped waveguide with both $40$ and $100$ subdomains.

\begin{figure}[t]
\centering
\begin{minipage}[b]{0.47\textwidth}
  \centering
  \includegraphics[width=\linewidth]{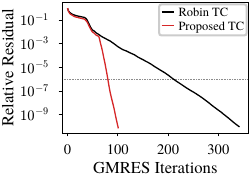}
  \convpanelmark{a}
\end{minipage}\hspace{0.04\textwidth}%
\begin{minipage}[b]{0.47\textwidth}
  \centering
  \includegraphics[width=\linewidth]{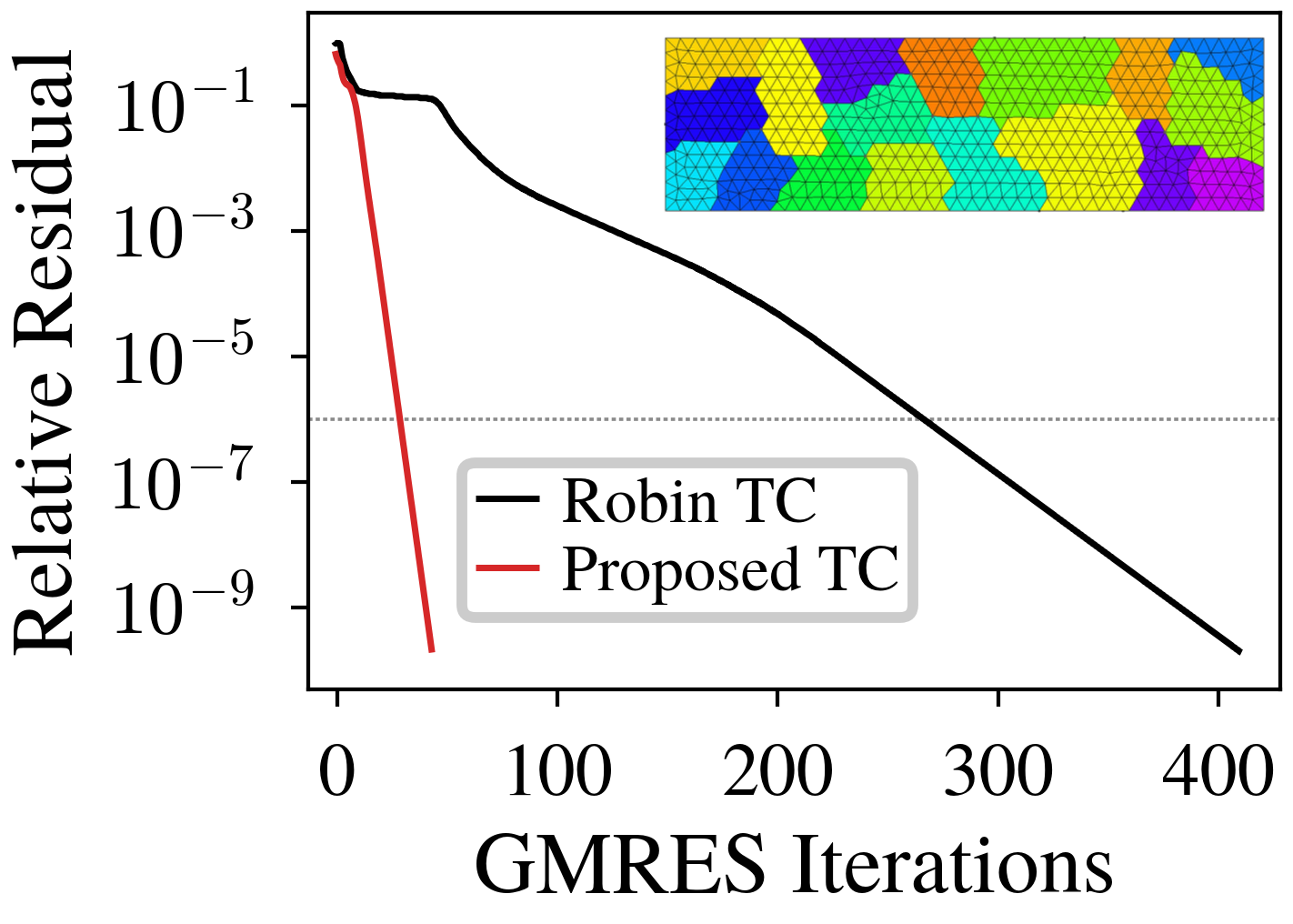}
  \convpanelmark{b}
\end{minipage}
\caption{GMRES convergence history of the global interface equation for ridge-loaded (a) and METIS-partitioned (b) waveguides. The inset in (b) shows the METIS mesh partition.}
\label{fig:wgs}
\end{figure}

\subsection{X-Band Bandpass Waveguide Filter}
\label{sec:filter}

\begin{figure}[t]
\centering
\subfloat[]{\includegraphics[width=0.8\textwidth]{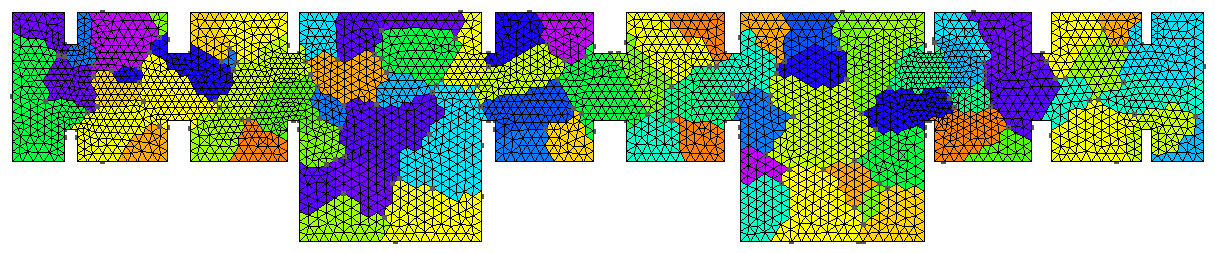}}\\[4pt]
\subfloat[]{\includegraphics[width=0.8\textwidth]{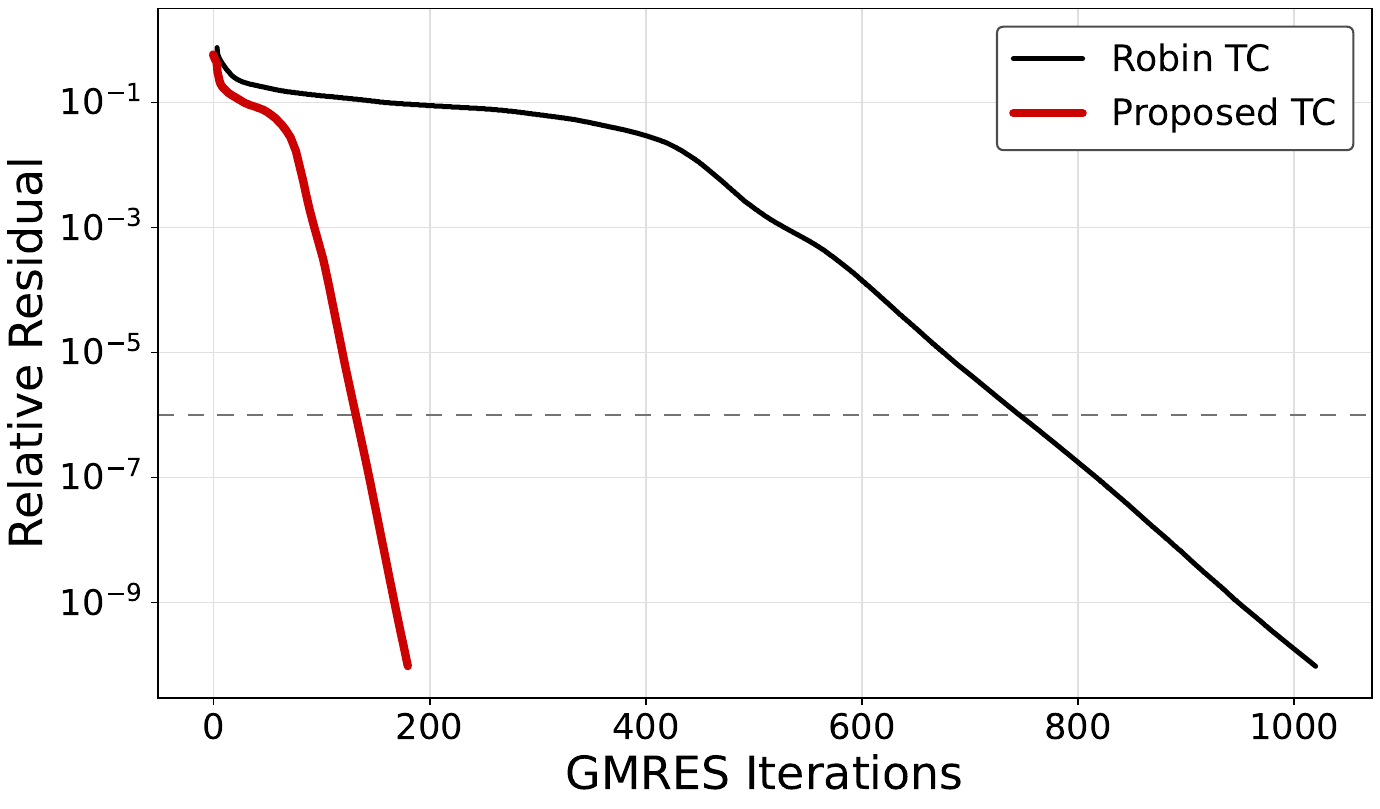}}
\caption{X-band bandpass waveguide filter, METIS-partitioned into $110$ subdomains: (a) partition geometry; (b) GMRES convergence history of the global interface equation.}
\label{fig:filter}
\end{figure}

A further example is an X-band bandpass waveguide filter for high-power SatCom applications, simulated at $8.4$~GHz. Its multiscale features, strong evanescent near-field coupling, and, once torn by METIS into $110$ subdomains, numerous corner edges make this a demanding problems for the Robin TC; the advantage of the proposed TC is particularly salient on this problem. Fig.~\ref{fig:filter}(a) shows the METIS-partitioned geometry using an
illustrative coarse mesh, whereas Fig.~\ref{fig:filter}(b) presents the corresponding GMRES convergence histories: the proposed TC reaches the $10^{-10}$ tolerance in 180 iterations against 1020 for the Robin TC, a nearly sixfold reduction. Consistent improvement was observed across all tested METIS partitionings with subdomain counts ranging from $50$ to $310$.

\subsection{Improved Interface Continuity}
\label{sec:jumps}

One of the principal advantages of the proposed two-channel FETI-DP-$4\lambda$ formulation, and one of the main drivers of the observed convergence acceleration, is the improvement in interface continuity over previous FETI approaches. Since the two channels directly impose continuity of tangential fields and of normal flux densities in the mixed $\mathbf E$-$\mathbf B$ setting, both continuities are enforced at every DDM interface. Consequently, significantly improved field and flux continuity is observed across all examples of this section. Table~\ref{tab:jumps} reports the measured jump statistics of $|\Delta(\hat{\bm n}\times\mathbf E)|$ and $|\Delta(\hat{\bm n}\cdot\mathbf B)|$ over all interface unknowns of the METIS-partitioned waveguide at the $10^{-6}$ stopping tolerance. At this tolerance, the worst-case normal-flux jump $|\Delta(\hat{\bm n}\cdot\mathbf B)|$ falls from $9.1\times10^{-7}$ under the Robin TC to $2.2\times10^{-12}$ under the proposed TC, a reduction of nearly six orders of magnitude, while the worst-case tangential-field jump $|\Delta(\hat{\bm n}\times\mathbf E)|$ decreases from $1.4\times10^{-8}$ to $1.2\times10^{-10}$, a reduction of more than two orders of magnitude. As the tolerance is tightened, both continuities improve, yet at every tolerance the proposed TC maintains a normal-flux jump orders of magnitude below that of the Robin TC. Beyond the faster convergence, the enhanced interface continuity improves total solution accuracy. At relaxed tolerance levels, e.g., $10^{-3}$, the large interface jumps of the Robin TC visibly degrade the solution, whereas the proposed TC yields a markedly more accurate solution.

\begin{table}[t]
\caption{Converged Interface Jump Statistics on the METIS-Partitioned Waveguide at
$10^{-6}$ Stopping Tolerance}
\label{tab:jumps}
\centering
\footnotesize
\setlength{\tabcolsep}{5pt}
\begin{tabular}{@{}lccc@{}}
\toprule
 & Mean & Median & Max \\
\midrule
\multicolumn{4}{@{}l}{$|\Delta(\hat{\bm n}\times\mathbf E)|$} \\
\quad Robin TC     & $3.3\times10^{-9}$ & $3.0\times10^{-9}$ & $1.4\times10^{-8}$ \\
\quad Proposed TC  & $\bm{1.4\times10^{-11}}$ & $\bm{4.1\times10^{-12}}$ & $\bm{1.2\times10^{-10}}$ \\
\midrule
\multicolumn{4}{@{}l}{$|\Delta(\hat{\bm n}\cdot\mathbf B)|$} \\
\quad Robin TC     & $7.7\times10^{-9}$ & $3.7\times10^{-10}$ & $9.1\times10^{-7}$ \\
\quad Proposed TC  & $\bm{1.2\times10^{-13}}$ & $\bm{5.4\times10^{-14}}$ & $\bm{2.2\times10^{-12}}$ \\
\bottomrule
\end{tabular}
\end{table}

\section{Conclusion}
In this letter, a novel two-channel transmission condition, constructed directly from the Faraday and Amp\`ere-Maxwell equations in a mixed $\mathbf{E}$-$\mathbf{B}$ setting, and the resulting FETI-DP-$4\lambda$ method were introduced for the DDM solution of large-scale EM problems. Across diverse 3-D scenarios spanning varying subdomain counts and volumes, partitionings, discretizations, multiscale features, and deep-evanescent waves, the proposed TC reduces the iteration counts of the global interface equation by up to several-fold relative to the widely used FETI approaches, while preserving the solution accuracy. Moreover, since both the tangential-field and normal-flux traces are enforced directly, the interface field and flux jumps decrease by orders of magnitude. Owing to the significantly improved convergence, the proposed TC enables robust, accurate, massively parallel, and scalable iterative DDM solvers for large-scale 3-D EM problems.

\bibliographystyle{IEEEtran}
\bibliography{references}

\end{document}